\documentstyle[11pt]{article}
 
\begin{document}

\title{CCD Photometry of SW Ursae Majoris during the 1996 Superoutburst}

\author{I.~~S~e~m~e~n~i~u~k,~~~A.~~O~l~e~c~h,~~~T.~~K~w~a~s~t~~~and~~~M.~~N~a~l~e~\.z~y~t~y}
\date{Warsaw University Observatory, Al.~Ujazdowskie~4,~00-478~Warszawa,
Poland\\
e-mail: (is,olech,tk,nalezyty)@sirius.astrouw.edu.pl}

\maketitle

\abstract{We present CCD $R$ photometry of SW~Ursae Majoris -- an SU~UMa type 
cataclysmic variable -- obtained during its April 1996 superoutburst. The mean 
value of the superhump period ${P_{\rm sh}}$ derived from our observations is 
0.05818($\pm2$)~days (83.8~min). The analysis of times of superhump maxima 
gives clear evidence for the increase of the superhump period with 
${\dot{P_{\rm sh}}=8.9\times10^{-5}}$. } 

\section{Introduction}

The SU~UMa type star, SW~Ursae Majoris, was discovered by Mrs L.\ Ceraski in 
1909 (Ceraski 1910), when as a result of an outburst it appeared over the 
threshold magnitude on an inspected plate. Due to its rare outbursts and their 
relatively great amplitudes Wellmann (1952) suggested that the star may be 
considered as an intermediate link between the U~Gem type stars and Nova-like 
or Nova stars. Discovery of the superhump phenomenon during the 1986 
superoutburst of the star (Robinson et al. 1987) made it clear that the object 
belongs to the SU~UMa type stars. 

The SU~UMa stars are a subclass of dwarf novae stars, whose distinctive 
feature is that they exhibit two essentially different types of outbursts. 
Beside the so called short or normal outbursts (normal maxima), typical for 
all dwarf novae and lasting a few days, the SU~UMa stars show the so called 
superoutbursts or supermaxima. The superoutbursts last 5--10 times longer than 
the normal outbursts, and their amplitudes exceed that of the normal outbursts 
by about 1~mag. A characteristic feature of the superoutbursts light curves is 
their sloping plateau which begins just after the supermaximum brightness and 
ends with a rapid decline to the minimum state. All SU~UMa type stars -- and 
this is their most distinctive defining mark -- show during superoutbursts 
some short period light variability, the so called superhumps, repeating with 
a period of the order of 100~min and amplitude of about 0.2--0.3~mag. The 
superhump periods of the SU~UMa stars are by a few percent longer than their 
orbital periods derived from spectroscopy or photometry obtained during 
quiescence. Some SU~UMa stars, like other dwarf novae, reveal the so called 
orbital humps during quiescence i.e. light modulation repeating with the 
orbital period that results from a favorable inclination of the binary orbit 
with respect to the observer. It should be stressed that superhumps appear on 
superoutburst light curves of all SU~UMa stars independently of whether they 
show orbital humps during quiescence or not, i.e. independently of 
inclination of their orbits. 

Recently Howell, Szkody and Cannizzo (1995) distinguished a subgroup of the 
SU~UMa type stars which they called "tremendous outburst amplitude dwarf 
novae" or "TOADs". Their superoutburst amplitudes are in the range 6--10~mag. 
At the same time the TOADs are the SU~UMa systems with the shortest orbital 
periods, the longest intervals between superoutbursts and they hardly ever 
undergo normal outbursts. For the SU~UMa stars of longer orbital periods and 
smaller superoutburst amplitudes normal outbursts are generally observed much 
more frequently than superoutbursts. SW~UMa with its superoutburst amplitude 
of about 7~mag was included into the TOADs. 

SW~Ursae Majoris is a star of ${V=16.5-17}$ at quiescence. Its orbital period 
determined from spectroscopy (Shafter, Szkody and Thorstensten 1986) and 
improved with the aid of photometric observations of orbital humps visible on 
the quiescence light curve of the star (Szkody, Osborne and Hassall 1988) is 
equal to 0.056815~days (81.8~min). It is the second shortest orbital period 
among the SU~UMa type stars (Warner 1995) with their orbital periods 
determined observationally.  

The mean recurrent time of SW~UMa superoutbursts, given by Howell, Szkody and 
Cannizzo (1995) is 400 days. According to Howell et al. (1995) in the time 
interval between November 1977 and August~1993 the AAVSO observers recorded 10 
outbursts of SW~UMa. Only one of these outbursts was a normal one. 

The superhump period of SW~UMa was determined as equal to 0.05833~days 
(84.0~min) by Robinson et al. (1987) during the 1986 superoutburst. To our 
knowledge the only other published observations of superhumps are those from 
the March 1992 superoutburst reported by Kato, Hirata and Mineshige (1992). 
The obtained then superhump period value was in good agreement with the 
Robinson et al. one. The superhump period of SW~UMa is by about 3\% longer than 
the orbital period. 
 
In April 1996 we were notified by the electronic {\it VSNET}, run by Drs T.
Kato and D. Nogami of the Kyoto University in Japan, that a superoutburst of 
SW~UMa had just begun. In the present paper we report on results of CCD 
photometry of SW~UMa performed during this superoutburst. 
   
\section{Observations}

The present superoutburst observations of SW~UMa were carried out at the 
Ostrowik station of the Warsaw University Observatory with a TK512~CCD at the 
Cassegrain focus of the 0.6~m telescope. The camera is described by Udalski 
and Pych (1992). We have monitored the star in the Cousins $R$ filter on 9 
nights from April 17 to 27, 1996. The exposure times were varied between 20 
and 45 seconds, depending on atmospheric conditions, dead time between the 
frames was 15 seconds. Journal of observations is given in Table~1. 

\vspace{7pt}
\begin{center}
Table 1\\
Journal of the Ostrowik CCD observations of SW~UMa\\
\vspace{7pt}

\begin{tabular}{|l|c|c||l|c|c|}
\hline
Date & Time of start & Length of & Date & Time of start & Length of  \\
1996 & JD~2450000.~+ & run (h)  & 1997 & JD~2450000.~+ & run (h)   \\
\hline
Apr 17    &  191.311 & 3.7 & Apr 22    &  196.268 & 7.0 \\
Apr 18    &  192.291 & 4.2 & Apr 23    &  197.314 & 2.8 \\
Apr 19    &  193.380 & 4.2 & Apr 24    &  198.299 & 1.2 \\
Apr 20    &  194.300 & 3.6 & Apr 27    &  201.348 & 4.9 \\
Apr 21    &  195.276 & 6.9 &           &          &     \\
\hline
\end{tabular}
\end{center}

The data reduction was performed using IRAF. The profile photometry was done 
with the DAOphotII package. The stars GSC~3798-0481 and GSC~3798-0491 served 
as the comparisons in obtaining relative $\Delta R$ magnitudes of SW~UMa. The 
first of these stars is a secondary photometric standard in the field of 
SW~UMa (Misselt 1996) with $R$ magnitude equal to 12.68, so we have eventually 
decided to reduce all the observations of SW~UMa to the observed instrumental 
$R$ magnitudes. 

According to the observations of the {\it VSNET} observers SW~UMa started its 
rapid superoutburst at the end of April 11 and attained the supermaximum 
brightness at the beginning of April 13. We started to monitor the star on 
April 17, i.e. about four days after the moment of maximum superoutburst 
brightness, when its brightness fell already by about 0.5~mag. Fig.~1 shows 
a general decrease of brightness of SW~UMa in the time interval of our 
observations, characteristic for the sloping plateau of superoutbursts of 
SU~UMa type stars. During 11 days the brightness fell by about 1~mag. 
 
Fig.~2 presents nightly light curves of SW~UMa for the nine April nights. 
First eight nights are consecutive, while the last one is separated from them 
by two nights. Superhumps, a typical phenomenon of the SU~UMa type stars 
superoutbursts, are observed on each night with their characteristic profile 
of steeper increase to the maximum and slower decrease. Their amplitude 
decreases during the eight consecutive nights from 0.2~mag on April 17 to 
about 0.07~mag on April 23 and 24. On the last night of our observations 
(April 27) the superhump amplitude increased slightly to about 0.12~mag and we 
observed a greater scatter of the observations resulting probably from a rapid 
flickering.   
 
\section{The Superhump Period}

From our observations we have determined 26 times of superhumps maxima. They 
are listed in Table~2. 

\vspace{7pt}
\begin{center}
Table 2\\
Times of Superhump Maxima of SW~UMa\\
\vspace{7pt}   
\begin{tabular}{|l|c|c||l|c|c|}
\hline
~HJD & $E$ & ${O-C}$  &~~~HJD & $E$ & ${O-C}$  \\
2450000.~+ &   & cycles  &2450000.~+  &  & cycles   \\  
\hline
&&&&&\\
   191.3345 &    0 &~~0.0870 & 195.5154 &   72 &--0.0559 \\
   191.3926 &    1 &~~0.0856 & 196.2770 &   85 &  0.0337 \\
   191.4488 &    2 &~~0.0515 & 196.3342 &   86 &  0.0168 \\
   192.3190 &   17 &~~0.0076 & 196.3943 &   87 &  0.0497 \\
   192.3766 &   18 &--0.0024 & 196.4520 &   88 &  0.0414 \\
   192.4360 &   19 &~~0.0185 & 196.5033 &   89 &--0.0769 \\
   193.4217 &   36 &--0.0403 & 197.3243 &  103 &  0.0336 \\
   193.4793 &   37 &--0.0503 & 197.3844 &  104 &  0.0666 \\
   193.5373 &   38 &--0.0535 & 198.3176 &  120 &  0.1055 \\
   194.3508 &   52 &--0.0719 &  \multicolumn{3}{|c|}{} \\
   194.4089 &   53 &--0.0733 &  \multicolumn{3}{|c|}{\raisebox{4pt}{Late Superhumps}}\\
   195.3418 &   69 &--0.0396 & 201.4057 & 172.5&  0.6806 \\
   195.3972 &   70 &--0.0874 & 201.4628 & 173.5&  0.6619 \\
   195.4578 &   71 &--0.0459 & 201.5214 & 174.5&  0.6691\\ 
\hline
\end{tabular}
\end{center}

In this Table only the first 23 times observed during the eight consecutive 
nights are the times of normal superhumps. The last three maxima are shifted 
by about half superhump period  compared with the earlier ones and we assume 
that they relate rather to the phenomenon of so called late superhumps, 
observed frequently at the end of superoutburst of SU~UMa type stars (Vogt 
1983, Udalski 1990, Warner 1995). Therefore, to determine the superhump 
period we have analyzed only the first 23 maxima of Table~2. The best linear 
fit to these maxima obtained with the least squares method gives the following 
ephemeris: 

$$\begin{tabular}{r@{\hspace{2pt}}c@{\hspace{2pt}}r@{\hspace{2pt}}c@{\hspace{2pt}}r@{\hspace{2pt}}l}
HJD$_{\rm Max}$ & = & 2450191.3294   &   +   & 0.05818 & $E$ \\
                &   & ${\pm}$~0.0014 & $\pm$ & 0.00002 &     \\
\end{tabular}
\eqno(1)$$

The obtained superhump period is slightly shorter than the value 0.05833~days 
determined by Robinson et al. (1987) from the 1986 superoutburst. Our value is 
a mean superhump period of the eight consecutive nights. The true superhump 
period is changing. This can be seen from the third and sixth columns of 
Table~2, where we have given the ${O-C}$ values calculated with the 
ephemeris~(1). These residuals evidently indicate an increase of the period. 
The quadratic ephemeris obtained as the best least squares fit to the same 23 
maxima as used previously is the following: 

$$\begin{tabular}{r@{\hspace{2pt}}c@{\hspace{2pt}}r@{\hspace{2pt}}c@{\hspace{2pt}}r@{\hspace{2pt}}l@{\hspace{2pt}}c@{\hspace{2pt}}@{\hspace{2pt}}l@{\hspace{2pt}}}
HJD$_{\rm Max}$ & = & 2450191.3339   &   +   & 0.05790 & $E$ & $+$   &
2.6$\times10^{-6}E^2$\\
                &   & ${\pm}$~0.0009 & $\pm$ & 0.00004 &     & $\pm$ & 0.3                  \\
\end{tabular}
\eqno(2)$$

The increase of the superhump period is shown in Fig.~3, where we have plotted 
the ${O-C}$ residuals taken from Table~2. The solid line in Fig.~3 presents 
the fit corresponding to the quadratic ephemeris~(2). The open circles 
correspond to the three last times of Table~2, which we interpret as the times 
of the late superhumps. The ${O-C}$ values for these times were calculated 
with the same ephemeris~(1) as the other residuals, only their cycle numbers 
$E$ were augmented by~0.5. 

\section{Interpulses and Late Superhumps} 

\vspace{7pt}
\begin{center}
Table 3\\
Times of Interpulse Maxima of SW~UMa\\
\vspace{7pt}
\begin{tabular}{|l|c|c||l|c|c|}
\hline
~HJD & $E$ & ${O-C}$  &~~~HJD & $E$ & ${O-C}$  \\
2450000.~+ &   & cycles  &2450000.~+  &  & cycles   \\  
\hline
&&&&&\\
   195.489 &  71.5 &--0.010 & 196.539 &  89.5 &  0.037 \\
   196.306 &  85.5 &  0.032 & 197.357 & 103.5 &  0.096 \\
   196.365 &  86.5 &  0.046 & & & \\
\hline
\end{tabular}
\end{center}

Beginning with the fifth night of our run (April 21) we can trace  secondary 
humps located on the light curves approximately midway between the main 
(normal) superhumps. In Fig.~2 they are marked with ticks. Such interpulses 
with progressively increasing amplitude were also observed during the 1986 
superoutburst of SW~UMa (Robinson et al. 1987) and during its 1992 
superoutburst (Kato, Hirata and Mineshige 1992). They were also observed in 
certain stages of superoutbursts of other SU~UMa type stars. Udalski (1990) 
observed them for SU~UMa itself. It was suggested (Schoembs and Vogt 1980, 
Warner 1995, p.~199) that the late superhumps in VW~Hyi may develop out of 
such interpulses. The times of interpulses determined from our light curves 
are collected in Table~3. Their average phase relative to the superhump 
maximum is~0.55. Assuming that in SW~UMa the interpulses also represent the 
late superhumps that have just begun to emerge, we attempted to fit a 
quadratic ephemeris to the times of Table~3 supplemented with three last times 
from Table~2. The resulting quadratic term is equal to (${2.7\pm 
.4)\times10^{-6}}$, what, within the error limits, is the same as the 
quadratic term for the normal superhumps (ephemeris~(2)). This result may be 
considered as an evidence that the temporal evolution of the late superhump 
period follows the evolution of the normal superhump period. The ${O-C}$ 
residuals in Table~3 were calculated with the ephemeris~(1). 
  
\section{Conclusions}

 From our CCD $R$ photometry performed during the 1996 superoutburst of SW~UMa 
we have determined a mean value of its superhump period ${P_{\rm sh}}$ as 
being equal to 0.05818(2)~days. We have demonstrated that the instantaneous 
value of the superhump period increases during the superoutburst. The period 
derivative ${\dot{P_{\rm sh}}}$, obtained from a parabolic fit to the ${O-C}$ 
diagram, is equal to ${8.9\times10^{-5}}$. Such a rate of the superhump period 
change is typical of SU~UMa stars. Generally, however, the SU~UMa stars show 
decreasing superhump periods during the declining plateau phase of 
superoutbursts. All SU~UMa stars with measured superhump period changes listed 
by Warner(1985) and Patterson et al. (1993) have negative ${\dot{P_{\rm sh}}}$. 
Hence, SW~UMa would be, to our knowledge, the first SU~UMa type star with well 
proved superhump period increase. Recently, Howell et al. (1996) and Patterson 
et al. (1996) published ${O-C}$ diagrams for the superhump period of AL~Comae 
Berenices observed during the 1995 superoutburst of the object. The diagrams 
seem to suggest that at least in certain time interval during the 
superoutburst of AL~Com the superhump period derivative was positive. Like 
SW~UMa, AL~Com is also a member of the TOADs (Howell, Szkody and Cannizzo 
1995). It might suggest that the SU~UMa stars with the shortest orbital 
periods and the longest superoutburst recurrence times would have preferably 
increasing superhump periods contrary to the other SU~UMa stars whose 
${\dot{P_{\rm sh}}}$ have negative values. 

Under the assumption that interpulses (secondary humps), visible on the light 
curves of SW~UMa beginning from a middle phase of the superoutburst, developed 
into the late superhumps observed on the night of April 27, we have shown that 
times of their maxima evolve in the same manner as those of the normal 
superhumps. The quadratic terms in parabolic fits describing the temporal 
behavior both of the normal superhump times as well as interpulses 
supplemented with late superhumps are, within the error limits, the same. It 
may be an evidence that interpulses and late superhumps are related to the 
same physical mechanism which underly normal superhumps and makes  both 
phenomena to appear in antiphase. 

Kato, Hirata and Mineshige (1992) reported the discovery of quasi-periodic 
oscillations with a recurrence time of about 6.1~min visible during the early 
phase of the 1992 superoutburst. We have undertaken an attempt to search for 
some other periodicities in our observational material. We have fitted, for 
each night independently, the superhump profiles with a combination of three 
sinusoids with frequencies corresponding to the superhump period and to its 
first and second overtones. We have removed this fit from observations and 
performed periodogram analysis for the obtained residuals. In resulting 
periodograms for the eight consecutive nights only the fourth and fifth 
harmonics of the superhump period appeared with some significant power. 
A rather week peak around the frequency corresponding to about 6~min could be 
seen only in the periodogram for the night of April~27.\\
 
\noindent {\bf Acknowledgments} We wish to express our gratitude to Professor Andrzej Kruszewski for  
having given us his observational time, for making for us the  observations 
from April 20, for many helpful discussions and comments  on the manuscript.\\ 
This work was partly supported by the KBN grant BW to the Warsaw University 
Observatory.

\vspace{0.5cm}
\begin{center}
REFERENCES
\end{center}
\vspace{0.4cm}

\noindent{Ceraski, W.},~{1910},~{Astron.~Nachr.},~{109},~{109}

\noindent{Howell, S.B., DeYoung, J.A., Mattei, J.A., Foster, G., Szkody, P.
Cannizzo, J.K., Walker, G., and Fierce, E.},~{1996},~{AJ},~{111},~{2367}

\noindent{Howell, S.B., Szkody, P., and Cannizzo, J.K.},~{1995},~{ApJ},~{439},~{337}

\noindent{Howell, S.B., Szkody, P., Sonneborn, G., Fried, R., Mattei, J. 
Oliversen, R.J., Ingram, D. and Hurst, G.M.},~{1995},~{ApJ},~{453},~{454}

\noindent{Kato, T., Hirata, R., and Mineshige, S},~{1992},~{PASJ},~{44},~{L215}

\noindent{Misselt, K.A.},~{1996},~{PASP},~{108},~{146}

\noindent{Patterson, J., Augusteijn, T., Harvey, D.A., Skillman, D.R., 
Abbott, T.M.C., and Thorstensen, J.},~{1996},~{PASP},~{108},~{748}

\noindent{Patterson, J., Bond, H.E., Grauer, A.D., Shafter, A.W., and
Mattei, J.A.},~{1993},~{PASP},~{105},~{69}

\noindent{Robinson, E.L., Shafter, A.W., Hill, J.A., Matt, A.W., and
Mattei, J.A.},~{1987},~{ApJ},~{313},~{772}

\noindent{Schoembs, R., and Vogt, N.},~{1980},~{A\&A},~{91},~{25}

\noindent{Shafter, A.W., Szkody, P., and Thorstensten, J.R.},~{1986},~{ApJ}
{308},~{765}

\noindent{Szkody, P., Osborne, J. and Hassall, B.J.M.},~{1988},~{ApJ},~{328},~{243}

\noindent{Udalski, A.},~{1990},~{AJ},~{100},~{226}

\noindent{Udalski, A. and Pych, W.},~{1992},~{Acta Astron.},~{42},~{285}

\noindent{Vogt, N.},~{1983},~{A\&A},~{118},~{95}

\noindent{Wellmann, P.},~{1952},~{Zs.~Ap.},~{31},~{123}

\noindent{Warner, B.},~{1985},~{"Interacting Binaries", ed. P.P. Eggleton and 
J.E. Pringle (Dodrecht, Reidel) 367}

\noindent{Warner, B.},~{1995},~{"Cataclysmic Variable Stars",
(Cambridge), Cambridge University Series 28}

\vspace{1cm}
\begin{center}
FIGURE CAPTION
\end{center}
\vspace{0.5cm}

\begin{itemize}
 
\item[Fig. 1.] The general decrease of brightness of SW Ursae Majoris 
        observed during the plateau phase of the 1996 superoutburst.
      
\item[Fig. 2.] The light curves of SW Ursae Majoris observed during 
        nine nights of April 1996. Ticks in the panels for the nights 
        of April 21, 22 and 23 indicate positions of interpulses.
             
\item[Fig. 3.] The O - C values for times of superhump maxima calculated
        with the linear ephemeris (1). The open circles correspond to 
        the three last values of Table~2, which we interpret as late
        superhump times. Their cycle numbers were augmented by 0.5. The 
        solid line presents the fit corresponding to the quadratic 
        ephemeris (2).
        
\end{itemize}

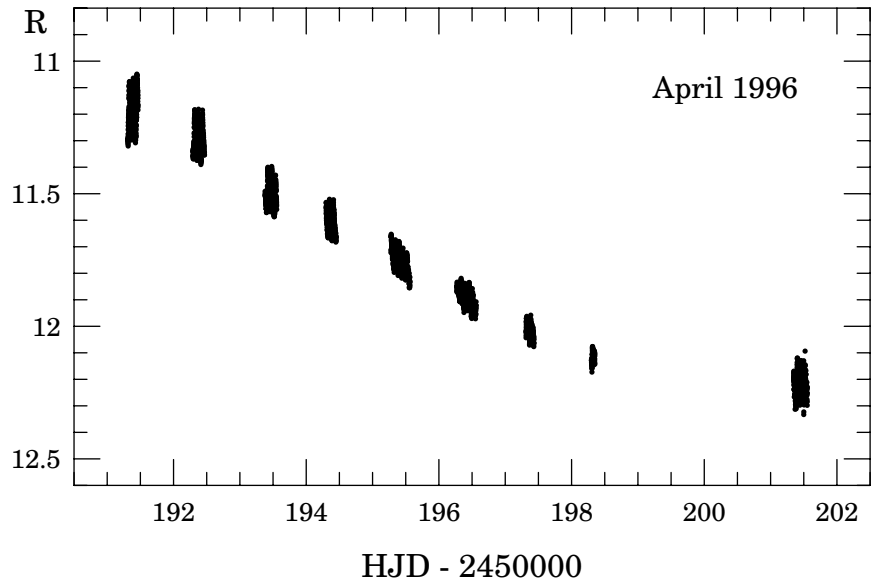
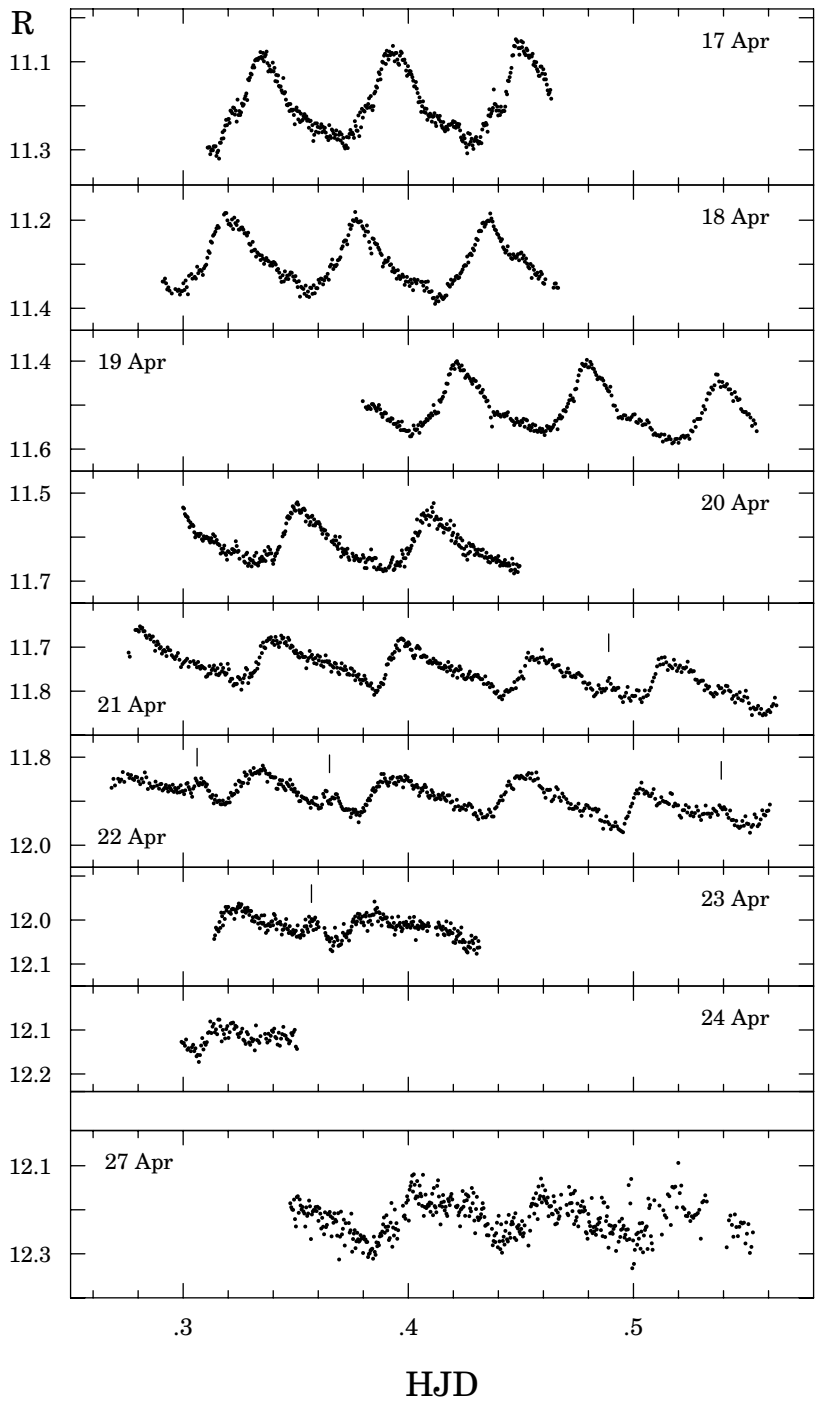
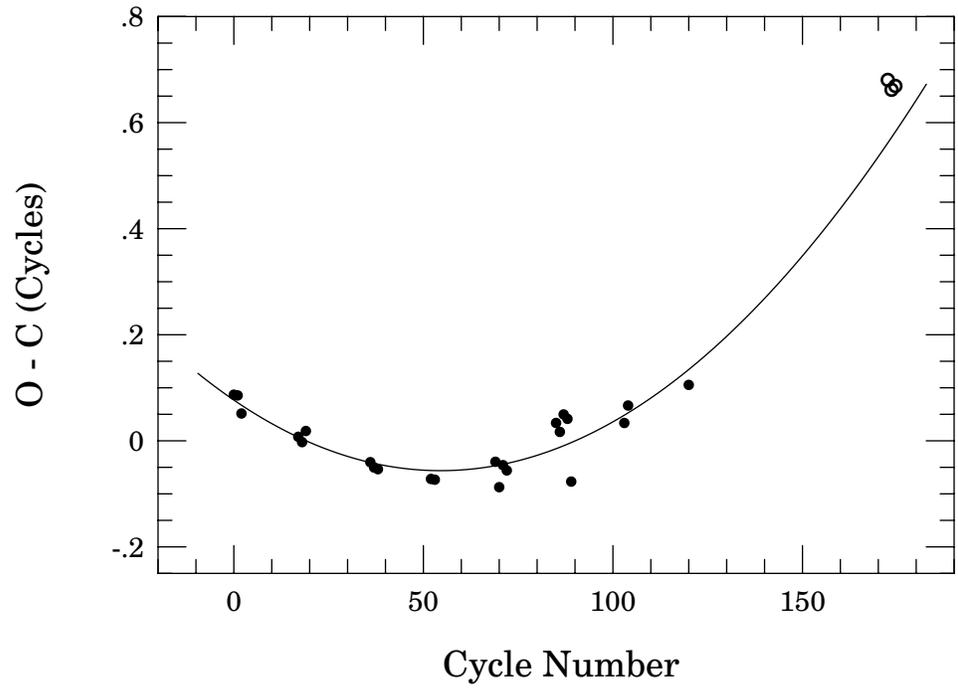
\end{document}